\documentclass[10pt,showpacs,aps,nofootinbib,superscriptaddress,twocolumn,longbibliography,prd]{revtex4}  %superscriptaddress nofootinbib
\usepackage{mathrsfs,amsmath,amsthm,latexsym,amssymb,amsfonts,epsfig,dsfont,txfonts,cancel}
\newcommand{\makeSymbol}[1]{\mathord{\vcenter{\hbox{#1}}}}
\usepackage{hyperref}
\hypersetup{bookmarks=true, bookmarksnumbered=true,
bookmarksopen=true, colorlinks, linkcolor=red, citecolor=magenta,
plainpages=false, pdfstartview=FitH}
\allowdisplaybreaks
\begin{document}

\title{New volume and inverse volume operators for loop quantum gravity}
\author{Jinsong Yang}\thanks{yangksong@gmail.com}
\affiliation{Department of Physics, Guizhou university, Guiyang 550025, China}
\affiliation{Institute of Physics, Academia Sinica, Taipei 115, Taiwan}
\author{Yongge Ma}\thanks{
Corresponding author. mayg@bnu.edu.cn}
\affiliation{
Department of Physics, Beijing Normal University, Beijing 100875, China}

\begin{abstract}
A new alternative volume operator is constructed for loop quantum gravity by using the so-called cotriad operators as building blocks. It is shown that the new volume operator shares the same qualitative properties with the standard volume operator. Moreover, a new alternative inverse volume operator is also constructed in the light of the construction of the alternative volume operator, which is possessed of the same qualitative properties as those of the alternative volume operator. The new inverse volume operator can be employed to construct the Hamiltonian operator of matter fields, which may lead to an anomaly-free on-shell quantum constraint algebra without any special restriction on the regularization procedure for gravity coupled to matter fields.

\end{abstract}

\maketitle

\section{Introduction}

To quantize a classical quantity, regularization procedures are often adopted. However, in general, different regularizations lead to different operators, and hence the so-called quantum ambiguities arise. In loop quantum gravity (LQG)  (see \cite{Ashtekar:2004eh,Han:2005km} for review articles, and \cite{Rovelli:2004tv,Thiemann:2007bk} for books), on one hand, the flux $\tilde{E}_i(S)$, as a classical basic variable of LQG, corresponds directly to the standard momentum operator. On the other hand, it can also be regularized as an alternative operator using the cotriad operator $\hat{e}_I$ corresponding to the integration of cotriad $e_a$ over a one-dimensional segment $s_I$. A celebrated result is that the standard and the alternative flux operators share the same properties, and they may become the same operator in a suitable setting \cite{Giesel:2005bk,Giesel:2005bm}.  As the cotriad operator enters the construction of the Hamiltonian constraint operator by Thiemann's trick and provides LQG a well-defined quantum dynamics \cite{Thiemann:1996aw}, the consistency check on the two flux operators enhances the confidence in the regularization procedure by using the cotriad operator as a building block to construct the quantum dynamics. In order to obtain the on-shell anomaly-free quantum constraint algebra, one has to employ degenerate triangulation at the coplanar vertices of spin networks in the regularization procedure of Thiemann's Hamiltonian. This problem has been overcome by a new proposed Hamiltonian constraint operator in \cite{Yang:2015zda}. In the general case of matter fields coupled to gravity, an inverse volume operator enters into the construction of the matter Hamiltonian operator \cite{Thiemann:1997rt}. In order to obtain also the on-shell anomaly-free quantum constraint algebra for the coupling system, the degenerate triangulation at coplanar vertices is still needed according to the treatment in \cite{Brunnemann:2005ip} for the original inverse volume operator. On the other hand, the inverse volume corresponds to the inverse scale factor in the isotropic cosmological model. In loop quantum cosmological models, the inverse scale factor operator is bounded above \cite{Ashtekar:2003hd}. This property is sometimes used to understand the big bang singularity resolution by loop quantum cosmology (LQC) \cite{Ashtekar:2003hd}. However, it is argued in \cite{Brunnemann:2005ip} that the boundedness of the inverse scale factor operator in LQC is not maintained by the inverse volume operator in LQG.

In this paper, the cotriad operator is used to construct new alternative volume and inverse volume operators. The graphical method introduced in \cite{graph-I,graph-II} brings much convenience to study the properties of these two operators in detail (see also \cite{DePietri:1996tvo,DePietri:1996pj} for the graphical method based on Tempereley-Lieb algebra). The new alternative volume operator shares the same properties with the standard volume operator in LQG \cite{Ashtekar:1997fb}. In the concrete, it is internal gauge invariant, diffeomorphism covariant, and symmetric. Its action on spin-network states leaves the spins $\vec{j}$ invariant but changes the intertwiners $\vec{i}$. It does not act at coplanar vertices or gauge-invariant trivalent vertices. The new alternative inverse volume operator is also possessed of the above properties. Thus we can employ this new inverse volume operator in the construction of matter Hamiltonian operators in the matter coupling cases. Since it does not act at coplanar vertices automatically, we need not employ degenerate triangulation any more in the regularization procedure of the whole Hamiltonian constraint operator for an on-shell anomaly-free constraint algebra. Moreover, the fact that the new alternative inverse volume operator is bounded from above on the special and typical eigenstates of the standard volume operator with zero eigenvalue opens a possibility to lift the result of singularity resolution of LQC to LQG.

In what follows, we briefly recall the elements of LQG and establish our notations and conventions. The Hamiltonian formalism of general relativity (GR) is formulated on a four-dimensional manifold $M\cong\mathrm{R}\times \Sigma$, with $\Sigma$ being a three-dimensional manifold of arbitrary topology. In Ashtekar-Barbero variables $(A_a^i,\tilde{E}^a_i)$ [indices $a,b,c,...$ refer to the tangent space of $\Sigma$  and $i,j,k,...$ to the $su(2)$ Lie algebra], GR can be cast as a dynamical theory of $SU(2)$ connections. The phase space is determined by
\begin{align}
\{A^i_a(x),\tilde{E}^b_j(y)\}=\kappa\beta\delta^b_a\delta^i_j\delta^3(x,y)\,,
\end{align}
where $\kappa=8\pi G$, and $\beta$ is the Barbero-Immirzi parameter. By $\gamma$ we denote a closed, piecewise analytic graph embedded in $\Sigma$, which is a set of edges that intersects at most at their end points. The collection of all end points of the edges in $\gamma$ is denoted by $V(\gamma)$, while the set of all edges in $\gamma$ is denoted by $E(\gamma)$. To construct quantum kinematics, one has to extend the configuration space ${\cal A}$ of smooth connections to the space $\bar{\cal A}$ of distributional connections. The projective techniques admit us to equip $\bar{\cal A}$ with a natural, faithful measure $\mu_o$, called the Ashtekar-Isham-Lewandowski measure, and then
the kinematical Hilbert space is given by  ${\cal H}_{\rm gr}:=L^2(\bar{\cal A},{\rm d}\mu_o)$. The bases of ${\cal H}_{\rm gr}$ are the so-called spin-network states $T_{\gamma,\vec{j},\vec{i}}(A):={\rm tr}\left(\otimes_{v\in V(\gamma)}\left[i_v\cdot\otimes_{e\in E(\gamma),v\in e}\pi_{j_e}(h_e(A))\right]\right)$, where $\vec{j}\equiv\{j_e\}_{e\in E(\gamma)}$ are the spins labeling edges of $\gamma$, and $\vec{i}\equiv\{i_v\}_{v\in V(\gamma)}$ are the set of intertwiners associated to vertices $v\in V(\gamma)$.

\section{New alternative volume operator}
\subsection{New alternative volume operator from the cotriad operator}
The cotriad operator $\hat{e}_I$, as a quantum version of the integral $e_I=\int_{s_I} e_a$ of the cotriad $e_a=e^i_a\tau_i$ (here $\tau_i=-\frac{i}{2}\sigma_i$ with $\sigma_i$ being the Pauli matrices) along a ``short'' segment $s_I$, has been widely applied to quantize a lot of physically interesting functions. For example, it appears in the construction of the length operator \cite{Thiemann:1996at}, the Hamiltonian constraints of pure gravity, and gravity coupled with matter. In this section, the cotriad operator is used to construct a new alternative volume operator.

\begin{figure}[htbp]
\centering
\includegraphics[width=8cm]{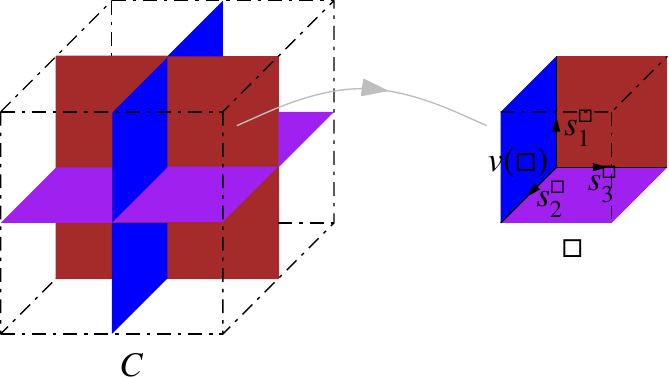}
\caption{An internal partition of cell $C$ into eight cubes $\Box$.}\label{figure-1}
\end{figure}

Classically, the volume function of an arbitrary three-dimensional region $R\subset\Sigma$ can be expressed as
\begin{align}\label{altv-classical}
V(R)&=\int_R{\rm d}^3x\,|\det(e^i_a)|=\int_R {\rm d}^3x\;\left|\frac{1}{3!}\tilde{\epsilon}^{abc}\epsilon_{ijk}e^i_ae^j_be^k_c\right|\notag\\
&=\int_R {\rm d}^3x\;\left|-\frac{4}{3!}\tilde{\epsilon}^{abc}{\rm tr}(e_ae_be_c)\right|\,,
\end{align}
where $e^i_a$ is the cotriad, and we have used $\epsilon_{ijk}=-4{\rm tr}(\tau_i\tau_j\tau_k)$. To regularize the volume function into a version that can be easily promoted as a well-defined operator in ${\cal H}_{\rm gr}$, we triangulate $R$ into a series of cells $C$ with parameter volume $\epsilon^3$ so that $\int_R=\sum\limits_{C\in R}\int_C$. The triangulation is denoted by $T(\epsilon)$. For each cell $C$, we choose an internal partition of $C$ into eight ``small" cubes incident to the central point of cell $C$. We call the central point of $C$ the base point of its eight cubes. Considering a cell $C$, we single out one cube $\Box$ from eight cubes, and denote $s_I(\Box), I=1,2,3$, the three edges of $\Box$ incident at the base-point $v(\Box)$ with outgoing direction (see Fig. \ref{figure-1}).
Then we obtain classically
\begin{align}\label{volume-reg}
V(R)&=\lim_{\epsilon\rightarrow0}\sum_{C\in T(\epsilon)}V_{C}=\lim_{\epsilon\rightarrow0}\sum_{C\in T(\epsilon)}\left|-\frac{4\times8}{3!}\epsilon^{IJK}{\rm tr}\left(e^\Box_Ie^\Box_Je^\Box_K\right)\right|\,,
\end{align}
where $\epsilon\rightarrow0$ corresponds to the process of shrinking $\Box$ to its base-point $v(\Box)$, and
\begin{align}
e^\Box_I:= -\frac{2}{\kappa\beta}h_{s_I(\Box)}\left\{h_{s_I(\Box)}^{-1},V\right\}
\end{align}
is the approximation of the integral of $e_a(x)=\frac{2}{\kappa\beta}\{A_a(x),V\}$ along $s_I(\Box)$. Here $V$ is the volume function of any region containing $x$.

To proceed from the classical formula \eqref{volume-reg} to the quantum expression, we first promote the classical holonomies and volume function to their quantum versions. Then we replace Poisson bracket by commutator times $1/(i\hbar)$. Thus the classical volume function $V(R)$ can be quantized as
\begin{align}\label{hat-e-def}
\hat{V}_{\rm alt}(R)=\lim_{\epsilon\rightarrow0}\sum_{C\in T(\epsilon)}\hat{V}^{\rm alt}_{C}=\lim_{\epsilon\rightarrow0}\sum_{C\in T(\epsilon)}\left|-\frac{4\times8}{3!}\epsilon^{IJK}{\rm tr}\left(\hat{e}^\Box_I\hat{e}^\Box_J\hat{e}^\Box_K\right)\right|\,,
\end{align}
with $\hat{e}^\Box_I:= -\frac{2}{i\beta\ell_{\rm p}^2}h_{s_I(\Box)}\left[h_{s_I(\Box)}^{-1},\hat{V}_{\rm std}\right]$, where $\ell_{\rm p}^2:=\hbar\kappa$, and $\hat{V}_{\rm std}$ denotes the $standard$ volume operator defined in \cite{Ashtekar:1997fb,Thiemann:1996au}.

Now we study the action of $\hat{V}^{\rm alt}_C$ on a cylindrical function $f_\gamma$ over a graph $\gamma$. The result depends on the relation between $C$ and $\gamma$ (or $\Box$ and $\gamma$). The relation contains the following three cases: (i) $s_I(\Box)\cap\gamma=\emptyset$; (ii) $s_I(\Box)\cap\gamma\neq\emptyset$ but does not contain a vertex of $\gamma$; (iii) $s_I(\Box)\cap\gamma\neq\emptyset$ and contains a vertex of $\gamma$. Detailed discussions parallel to those of \cite{Thiemann:1996aw} show that the results are trivial (or 0) in the former two cases. Hence it is natural for us to partition $R$ in the way adapted to $\gamma$. For a given $v\in V(\gamma)$, we single out a noncoplanar triple of edges of $\gamma$ at the vertex $v$, and denote the corresponding three segments of the triple of edges starting from $v$ with parameter length $\epsilon$ by $s_I,s_J,s_K$. We then construct a cube $\Box$ adapted to the three segments $s_I,s_J,s_K$ such that the three segments are the boundary edges of $\Box$. By reversely extending $s_I,s_J,s_K$ we obtain three more segments. Then we can get another seven ``mirror'' cubes with the six segments. Together with $\Box$, the eight cubes consist of the cell $C$ for which we look. Since the information of $\Box$ has been implied in the segments, we can omit the notation $\Box$ for simplification. Then Eq. \eqref{hat-e-def} reduces to
\begin{align}\label{trival-lim-vol}
\hat{V}_{\rm alt}(R)\cdot f_\gamma&=\lim_{\epsilon\rightarrow0}\sum_{v\in V(\gamma)\cap R}\hat{V}^{\rm alt}_v\cdot f_\gamma\notag\\
&=\lim_{\epsilon\rightarrow0}\sum_{v\in V(\gamma)\cap R}\left|-\frac{4\times8}{3!E(v)}\sum_{s_I\cap s_J\cap s_K=v}\epsilon^{IJK}{\rm tr}\left(\hat{e}_I\hat{e}_J\hat{e}_K\right)\right|\cdot f_\gamma\,,
\end{align}
where $E(v)$ is the number of the triples of noncoplanar edges of $\gamma$ at $v$, and the second sum is over these triples. The case for the coplanar vertices is discussed in the next subsection. Similar to the discussion in \cite{Thiemann:1996at}, it turns out that $\hat{V}^{\rm alt}_v\cdot f_\gamma$ in \eqref{trival-lim-vol} is actually independent of the parameter length $\epsilon$ of $s_I$. Hence the limitation in \eqref{trival-lim-vol} can be taken trivially, which yields the alternative volume operator
\begin{align}\label{alt-vol}
\hat{V}_{\rm alt}(R)\cdot f_\gamma&=\sum_{v\in V(\gamma)\cap R}\hat{V}^{\rm alt}_v\cdot f_\gamma\notag\\
&=\sum_{v\in V(\gamma)\cap R}\left|-\frac{4\times8}{3!E(v)}\sum_{s_I\cap s_J\cap s_K=v}\epsilon^{IJK}{\rm tr}\left(\hat{e}_I\hat{e}_J\hat{e}_K\right)\right|\cdot f_\gamma\,.
\end{align}

\subsection{Properties of the alternative volume operator}\label{II-B}
We have presented the derivation of a new alternative volume operator \eqref{alt-vol} by using the so-called cotriad operator $\hat{e}_I$. A natural question is whether the two volume operators, the alternative volume operator \eqref{alt-vol} and the standard operator in \cite{Ashtekar:1997fb,Thiemann:1996au}, have the same properties. To answer this question, let us first recall the formula of the standard volume operator and its key properties, then analyze the properties of the alternative one.

The standard volume operator is derived from two regularization procedures in \cite{Ashtekar:1997fb,Thiemann:1996au}. Its only regularization ambiguity $\kappa_o$ is fixed as $\kappa_o=1$ in \cite{Giesel:2005bk,Giesel:2005bm}. Its action reads
\begin{align}\label{std-volume-def}
&\hat{V}_{\rm std}(R)\cdot f_\gamma=\sum_{v\in V(\gamma)\cap R}\hat{V}^{\rm std}_v\cdot f_\gamma\notag\\
&=\ell_{\rm p}^3\,\beta^{\frac32}\sum_{v\in V(\gamma)\cap R}\sqrt{\left|\frac{i}{3!\times8\times4}\sum_{e_I\cap e_J \cap e_K=v}\varsigma(e_I,e_J,e_K)\;\hat{q}_{IJK}\right|}\,\cdot f_\gamma\notag\\
&=\ell_{\rm p}^3\,\beta^{\frac32}\sum_{v\in V(\gamma)\cap R}\sqrt{\left|\frac{i}{8\times4}\sum_{\substack{I<J<K\\e_I\cap e_J \cap e_K=v}}\varsigma(e_I,e_J,e_K)\;\hat{q}_{IJK}\right|}\,\cdot f_\gamma
\end{align}
where $\varsigma(e_I,e_J,e_K):={{\rm sgn}(\det(\dot{e}_I(0),\dot{e}_J(0),\dot{e}_K(0)))}$, and $\hat{q}_{IJK}:=-4i\epsilon_{ijk}J^i_{e_I}J^j_{e_J}J^k_{e_K}$. Here $J^i_{e_I}$ is the essential self-adjoint right-invariant vector field on the copy of SU(2) corresponding to the $I$th edge, which is defined by
 \begin{align}
 &J^i_{e_I}\cdot f_\gamma(h_{e_1},\cdots,h_{e_I},\cdots,h_{e_n})\notag\\
 &\hspace{2cm}:=-i\left.\frac{\rm d}{{\rm d}t}\right|_{t=0}f_\gamma(h_{e_1},\cdots,e^{t\tau_i}h_{e_I},\cdots,h_{e_n}).
 \end{align}
 The standard volume operator has the following key properties: (i) The operator is gauge invariant and diffeomorphism covariant; (ii) it leaves the spins $\vec{j}$ invariant, but just changes the intertwiners $\vec{i}$ of the spin-network function $T_{\gamma,\vec{j},\vec{i}}$ on which it acts; (iii) it is symmetric; (iv) its action on a coplanar vertices of $\gamma$ is trivial; (v) its action on the gauge-invariant trivalent vertices is also trivial. In what follows we show step by step that the above properties of the standard volume operator \eqref{std-volume-def} are preserved by the alternative volume operator \eqref{alt-vol}. Hence the two volume operators possess the same qualitative properties.

It is easy to see that the alternative volume operator \eqref{alt-vol} is gauge invariant, since under a gauge transformation, the cotriad operator $\hat{e}_I$ changes to $g(v)\hat{e}_Ig(v)^{-1}$, which can be easily derived from the  gauge transformations of the standard volume operator and the holonomies. The fact that the standard volume operator and holomomies transform covariantly with respect to diffeomorphism ensures that the alternative volume operator \eqref{alt-vol} is also diffeomorphism covariant.

Now we show that the alternative volume operator also has the property (ii) of the standard one. Consider a vertex $v\in V(\gamma)$ at which three edges $e_I,e_J,e_K$ of $\gamma$ are incident, and denote their three segments starting at $v$ by $s_I, s_J, s_K$ while denoting the remaining segments by $l_I,l_J,l_K$.  The result of applying ${\rm tr}(\hat{e}_I\hat{e}_J\hat{e}_K)$ to $T_{\gamma,\vec{j},\vec{i}}$ is a composition of spin-network states in which there exist terms depending on $\gamma\cup s_I\cup s_J\cup s_K$ with the spins of the edges of $\gamma$ differing from the considered three edges by $j_{s_I}\in\{j_{e_I}\pm1,j_{e_I}\}$ while $l_I,l_J,l_K$ are unchanged. However, for the cases $j_{s_I}=j_{e_I}\pm1$ there exists no contractor at the divalent vertex  $s_I\cap l_I$ to make such a state gauge invariant.

To see that the alternative volume operator is symmetric, we notice $\hat{V}_{\rm std}=\hat{V}^\dag_{\rm std}$ and the identity
\begin{align}
\overline{{(h_{s_I})^A}_B}={(h_{s_I}^{-1})^B}_A\,.
\end{align}
Therefore we have
\begin{align}
\left[{(\hat{e}_I)^A}_B\right]^\dag&=\left[-\frac{2}{i\beta\ell_{\rm p}^2}{(h_{s_I}[h_{s_I}^{-1},\hat{V}_{\rm std}])^A}_B\right]^\dag\notag\\
&=\frac{2}{i\beta\ell_{\rm p}^2}\left[\hat{V}_{\rm std}\delta^A_B-\overline{{(h_{s_I}^{-1})^C}_B}\hat{V}_{\rm std}\overline{{(h_{s_I})^A}_C}\right]\notag\\
&=\frac{2}{i\beta\ell_{\rm p}^2}{(h_{s_I}[h_{s_I}^{-1},\hat{V}_{\rm std}])^B}_A\notag\\
&=-{(\hat{e}_I)^B}_A\,.
\end{align}
Hence, for a given $v$, the second sum of the terms in \eqref{alt-vol} can be written as
\begin{align}\label{symmetric-form}
&\sum_{s_I\cap s_J\cap s_K=v}\epsilon^{IJK}{\rm tr}\left(\hat{e}_I\hat{e}_J\hat{e}_K\right)\notag\\
&=\frac12\sum_{s_I\cap s_J\cap s_K=v}\left[\epsilon^{IJK}{\rm tr}\left(\hat{e}_I\hat{e}_J\hat{e}_K\right)+\epsilon^{KJI}{\rm tr}\left(\hat{e}_K\hat{e}_J\hat{e}_I\right)\right]\notag\\
&=\frac12\sum_{s_I\cap s_J\cap s_K=v}\epsilon^{IJK}\left\{{\rm tr}\left(\hat{e}_I\hat{e}_J\hat{e}_K\right)+\left[{\rm tr}\left(\hat{e}_I\hat{e}_J\hat{e}_K\right)\right]^\dag\right\}\,,
\end{align}
which implies that the alternative volume operator \eqref{alt-vol} is symmetric.

Let us turn to prove that the alternative volume operator acts trivially on a coplanar vertices. Consider a coplanar vertex $v\in V(\gamma)$ at which the tangent vectors of edges of $\gamma$ incident span a two-dimensional surface $S_v$. In order to partition the region containing $v$ adapted to $\gamma$, we need to add a new segment that is transversal to $S_v$ at $v$. With the new additional segment, we get cubes based at $v$. The form of the volume operator $\hat{V}^{\rm alt}_v$ at the coplanar vertex $v$ is the same as the one in \eqref{alt-vol}, but one of triple segments $s_I,s_J,s_K$ is the new additional segment at $v$. Assume that $s_I$ is a segment of one edge of $\gamma$. Then the action of cotriad operator $\hat{e}_I$ associated with $s_I$ on cylindrical functions $f_\gamma$ with respect to $\gamma$ yields
\begin{align}
\hat{e}_I\cdot f_\gamma&=-\frac{2}{i\beta\ell_{\rm p}^2}h_{s_I}\left[h_{s_I}^{-1},\hat{V}_{\rm std}\right]\cdot f_\gamma\notag\\
&=-\frac{2}{i\beta\ell_{\rm p}^2}\left(\hat{V}^{\rm std}_v-h_{s_I}\hat{V}^{\rm std}_vh_{s_I}^{-1}\right)\cdot f_\gamma=0\,,
\end{align}
where in the last step we used one property of the standard volume operator that vanishes at the coplanar vertex $v$. Hence, for a coplanar vertex $v\in V(\gamma)$, the nontrivial result of $\hat{V}^{\rm alt}_v$ applying to $f_\gamma$ corresponds to the case in which the triples $s_I,s_J,s_K$ are all transversal to $S_v$. On the other hand, for the nontrivial case, $\hat{e}_I$ does not change the graph and the spins associated to the edges of the spin network, but only changes the intertwiners. The result of this action is independent of the  intersection character with edge $s_I$, i.e., $\hat{e}_I\cdot f_\gamma=\hat{e}_J\cdot f_\gamma$. Hence, for a coplanar vertex $v$, the cotriad operators $\hat{e}_I$ commute with each other. This means that the result of ${\rm tr}(\hat{e}_I\hat{e}_J\hat{e}_K)$ acting on $f_\gamma$ does not depend on the ordering of the three cotriad operators. Therefore we get $\sum_{s_I\cap s_J\cap s_K=v}\epsilon^{IJK}{\rm tr}(\hat{e}_I\hat{e}_J\hat{e}_K)\cdot f_\gamma=0$. This completes the proof that the alternative volume operator \eqref{alt-vol} acts trivially at coplanar vertices.

Finally, we consider the action of the alternative volume operator on the gauge-invariant cylindrical function over $\gamma$ with trivalent vertices. Let us focus on a noncoplanar trivalent vertex $v$ of $\gamma$ and denote the three edges starting from $v$ by $e_1,e_2,e_3$. Then one has $I,J,K\in\{1,2,3\}$ appearing in the expression \eqref{alt-vol} of the alternative volume operator at $v$. Notice that the intertwiner space for a gauge-invariant trivalent vertex $v$ is one dimensional, and the operator ${\rm tr}(\hat{e}_I\hat{e}_J\hat{e}_K)$ is gauge invariant. Hence ${\rm tr}(\hat{e}_I\hat{e}_J\hat{e}_K)$ changes the intertwiner of a gauge-invariant trivalent vertex into itself up to a constant. In other words, any spin-network state with gauge-invariant trivalent vertices is an eigenvector of ${\rm tr}(\hat{e}_I\hat{e}_J\hat{e}_K)$. Denote the spins of the three edges $e_1,e_2,e_3$ incident at $v$ by $j_1,j_2,j_3$. Then the part of the spin-network state $T^{v,s}_{\gamma,\vec{j},\vec{i}}(A)$ (the notation $s$ denotes the segments $s_I$) corresponding to $v$ can be expressed graphically as (see \cite{graph-I,graph-II} for an introduction to the graphical method in LQG)
\begin{align}\label{graph-snf-v-s}
T^{v,s}_{\gamma,\vec{j},\vec{i}}(A)&=\left(i_v\right)_{\,m_1m_2m_3}{[\pi_{j_1}(h_{s_1})]^{m_1}}_{\,l_1}{[\pi_{j_2}(h_{s_2})]^{m_2}}_{\,l_2}{[\pi_{j_3}(h_{s_3})]^{m_3}}_{\,l_3}\notag\\
&=(-1)^{2j_3}\makeSymbol{
\includegraphics[width=3cm]{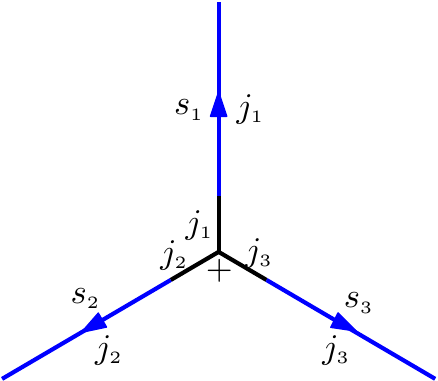}}\,,
\end{align}
with a normalized intertwiner
\begin{align}\label{tri-intertwiner}
(-1)^{2j_3}\makeSymbol{
\includegraphics[width=1cm]{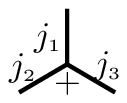}}&=\makeSymbol{
\includegraphics[width=2cm]{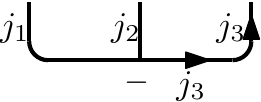}}\notag\\
&=\sqrt{2j_3+1}\makeSymbol{
\includegraphics[width=2.6cm]{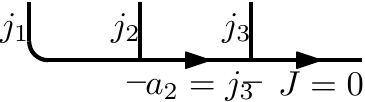}}\,.
\end{align}
Then at $v$, the action of $\hat{V}^{\rm alt}_v$ on $T^{v,s}_{\gamma,\vec{j},\vec{i}}(A)$ is
\begin{align}\label{alt-vol-trivalent}
&\hat{V}^{\rm alt}_v\cdot T^{v,s}_{\gamma,\vec{j},\vec{i}}(A)=\frac{16}{3}\left|\sum_{I,J,K\in\{1,2,3\}}\epsilon^{IJK}{\rm tr}\left(\hat{e}_I\hat{e}_J\hat{e}_K\right)\right|\cdot T^{v,s}_{\gamma,\vec{j},\vec{i}}(A)\notag\\
&=\frac{8}{3}\left|\sum_{I,J,K\in\{1,2,3\}}\epsilon^{IJK}\left\{{\rm tr}\left(\hat{e}_I\hat{e}_J\hat{e}_K\right)+\left[{\rm tr}\left(\hat{e}_I\hat{e}_J\hat{e}_K\right)\right]^\dag\right\}\right|\cdot T^{v,s}_{\gamma,\vec{j},\vec{i}}(A)\,,
\end{align}
where we have used Eq. \eqref{symmetric-form} in the second step. Our ultimate goal is to show that the result of \eqref{alt-vol-trivalent} vanishes. It would be enough if we could prove that ${\rm tr}\left(\hat{e}_I\hat{e}_J\hat{e}_K\right)$ takes a pure imaginary eigenvalue on its eigenstate $T^{v,s}_{\gamma,\vec{j},\vec{i}}(A)$ for given $I,J,K$. With the trivalent intertwiner \eqref{tri-intertwiner} as eigenvector, ${\rm tr}(\hat{e}_I\hat{e}_J\hat{e}_K)$ has the following eigenvalue up to a (real) normalized factor
\begin{widetext}
\begin{align}\label{eigenvalue-matrix}
\left\langle T^{v,s}_{\gamma,\vec{j},\vec{i}}(A)\right|{\rm tr}(\hat{e}_I\hat{e}_J\hat{e}_K)\left|T^{v,s}_{\gamma,\vec{j},\vec{i}}(A)\right\rangle
=&\left(-\frac{2}{i\beta\ell_{\rm p}^2}\right)^3\left\langle T^{v,s}_{\gamma,\vec{j},\vec{i}}(A)\right|{\rm tr}\left(h_{s_I}\hat{V}^{\rm std}_vh_{s_I}^{-1}\hat{V}^{\rm std}_vh_{s_K}\hat{V}^{\rm std}_vh_{s_K}^{-1}-h_{s_I}\hat{V}^{\rm std}_vh_{s_I}^{-1}h_{s_J}\hat{V}^{\rm std}_vh_{s_J}^{-1}h_{s_K}\hat{V}^{\rm std}_vh_{s_K}^{-1}\right)\left|T^{v,s}_{\gamma,\vec{j},\vec{i}}(A)\right\rangle\notag\\
=&\left(-\frac{2}{i\beta\ell_{\rm p}^2}\right)^3\sum_{A,C}\left\langle T^{v,s}_{\gamma,\vec{j},\vec{i}}(A)\right|\left[{(h_{s_I}\hat{V}^{\rm std}_vh_{s_I}^{-1})^A}_C\right]^\dag\hat{V}^{\rm std}_v{(h_{s_K}\hat{V}^{\rm std}_vh_{s_K}^{-1})^A}_C\left|T^{v,s}_{\gamma,\vec{j},\vec{i}}(A)\right\rangle\notag\\
&\qquad-\left(-\frac{2}{i\beta\ell_{\rm p}^2}\right)^3\sum_{B,C}\left\langle T^{v,s}_{\gamma,\vec{j},\vec{i}}(A)\right|\left[{(h_{s_J}^{-1}h_{s_I}\hat{V}^{\rm std}_vh_{s_I}^{-1})^B}_C\right]^\dag\hat{V}^{\rm std}_v{(h_{s_J}^{-1}h_{s_K}\hat{V}^{\rm std}_vh_{s_K}^{-1})^B}_C\left|T^{v,s}_{\gamma,\vec{j},\vec{i}}(A)\right\rangle\notag\\
=&\left(-\frac{2}{i\beta\ell_{\rm p}^2}\right)^3\sum_{A,C}\left\langle {\left(h_{s_I}\hat{V}^{\rm std}_vh_{s_I}^{-1}\right)^A}_C\cdot T^{v,s}_{\gamma,\vec{j},\vec{i}}(A)\right|\hat{V}^{\rm std}_v\left|{\left(h_{s_K}\hat{V}^{\rm std}_vh_{s_K}^{-1}\right)^A}_C\cdot T^{v,s}_{\gamma,\vec{j},\vec{i}}(A)\right\rangle\notag\\
&\qquad-\left(-\frac{2}{i\beta\ell_{\rm p}^2}\right)^3\sum_{B,C}\left\langle {\left(h_{s_J}^{-1}h_{s_I}\hat{V}^{\rm std}_vh_{s_I}^{-1}\right)^B}_C\cdot T^{v,s}_{\gamma,\vec{j},\vec{i}}(A)\right|\hat{V}^{\rm std}_v\left|{\left(h_{s_J}^{-1}h_{s_K}\hat{V}^{\rm std}_vh_{s_K}^{-1}\right)^B}_C\cdot T^{v,s}_{\gamma,\vec{j},\vec{i}}(A)\right\rangle\,,
\end{align}
\end{widetext}
where in the first step we used the fact that those terms where $\hat{V}^{\rm std}_v$ stands on the far left or right vanish because of the properties (iv) of $\hat{V}^{\rm std}_v$, and in the second step we have used
\begin{align}
\left[{(h_{s_I}\hat{V}^{\rm std}_vh_{s_I}^{-1})^A}_C\right]^\dag&=\overline{{(h_{s_I}^{-1})^B}_C}\hat{V}^{\rm std}_v\overline{{(h_{s_I})^A}_B}\notag\\
&={(h_{s_I}\hat{V}^{\rm std}_vh_{s_I}^{-1})^C}_A\,,
\end{align}
and
\begin{align}
\left[{(h_{s_J}^{-1}h_{s_I}\hat{V}^{\rm std}_vh_{s_I}^{-1})^B}_C\right]^\dag&=\overline{{(h_{s_I}^{-1})^E}_C}\hat{V}^{\rm std}_v\overline{{(h_{s_I})^D}_E}\overline{{(h_{s_J}^{-1})^B}_D}\notag\\
&={(h_{s_I}\hat{V}^{\rm std}_vh_{s_I}^{-1}h_{s_J})^C}_B\,.
\end{align}
Hence, to show that $\left\langle T^{v,s}_{\gamma,\vec{j},\vec{i}}(A)\right|{\rm tr}(\hat{e}_I\hat{e}_J\hat{e}_K)\left|T^{v,s}_{\gamma,\vec{j},\vec{i}}(A)\right\rangle$ takes a pure imaginary number is equal to proving that the two matrix elements of $\hat{V}^{\rm std}_v$ in Eq. \eqref{eigenvalue-matrix} are real. We consider the special case that $I=1,J=2,K=3$. It is easy to see that the symmetry ensures that the result for the special case still holds for the remaining cases. Direct calculations in Appendix \ref{matrix-elements} show that the two matrix elements of $\hat{V}^{\rm std}_v$ in Eq. \eqref{eigenvalue-matrix} can be calculated as
\begin{align}\label{first-term}
&\sum_{A,C}\left\langle{\left(h_{s_1}\hat{V}^{\rm std}_vh_{s_1}^{-1}\right)^A}_C\cdot T^{v,s}_{\gamma,\vec{j},\vec{i}}(A)\right|\hat{V}^{\rm std}_v\left|{\left(h_{s_3}\hat{V}^{\rm std}_vh_{s_3}^{-1}\right)^A}_C\cdot T^{v,s}_{\gamma,\vec{j},\vec{i}}(A)\right\rangle\notag\\
&=\sum_{\substack{A,C\\a'_2,a''_2,a'_3}}A(j_1,j_2,j_3;a'_2)B(j_1,j_2,j_3;a''_2,a'_3)\notag\\
&\hspace{1cm}\times\left\langle\makeSymbol{
\includegraphics[width=3.3cm]{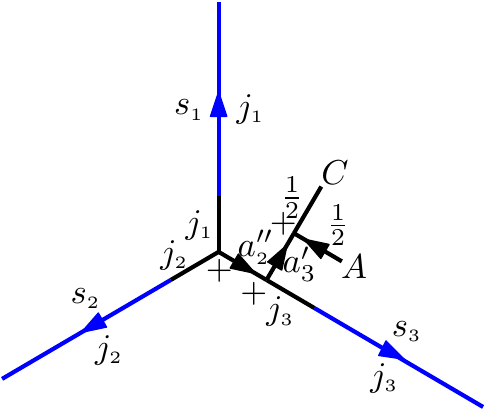}}\right|\left.\makeSymbol{
\includegraphics[width=3.3cm]{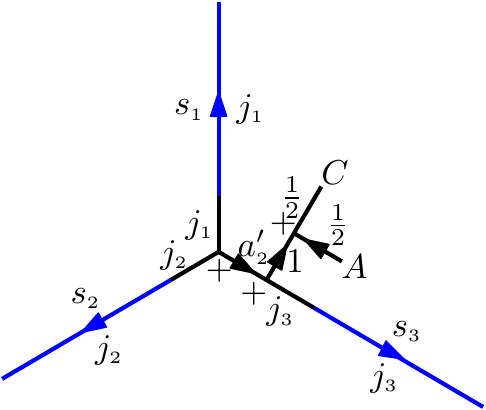}}\right\rangle\,,
\end{align}
and
\begin{align}\label{second-term}
&\sum_{B,C}\left\langle{\left(h_{s_2}^{-1}h_{s_1}\hat{V}^{\rm std}_vh_{s_1}^{-1}\right)^B}_C\cdot T^{v,s}_{\gamma,\vec{j},\vec{i}}(A)\right|\hat{V}^{\rm std}_v\left|{\left(h_{s_2}^{-1}h_{s_3}\hat{V}^{\rm std}_vh_{s_3}^{-1}\right)^B}_C\cdot T^{v,s}_{\gamma,\vec{j},\vec{i}}(A)\right\rangle\notag\\
=&\sum_{\substack{B,C\\j'_2,j''_2,a_2,a'_2}}D(j_1,j_2,j_3;j'_2;a_2)E(j_1,j_2,j_3;j''_2;a'_2)\notag\\
&\times\left\langle\makeSymbol{
\includegraphics[width=3.6cm]{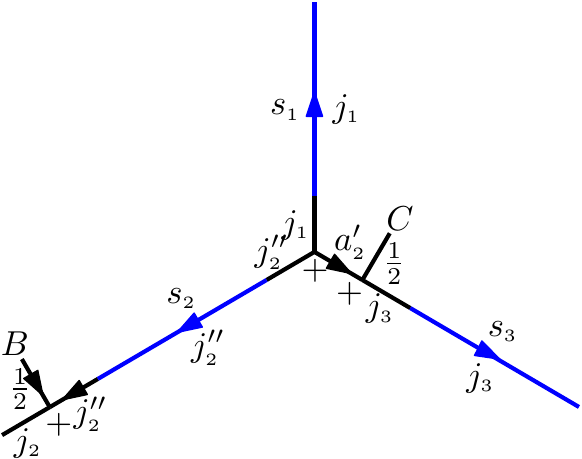}}\right|\left.\makeSymbol{
\includegraphics[width=3.6cm]{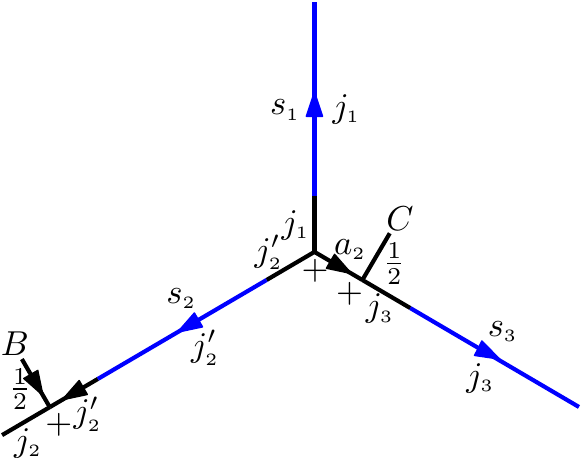}}\right\rangle\,,
\end{align}
where the factors are defined as
\begin{widetext}
\begin{align}\label{A-volume-matrix}
A(j_1,j_2,j_3;a'_2) &:=\sum_{j'_3}V(j'_3,j_1,j_2)(-1)^{2j_3}(2j'_3+1)3(-1)^{\frac12+a'_2-j'_3}\begin{Bmatrix}
\frac12 & \frac12 & 1\\
j_3 & j_3 & j'_3
\end{Bmatrix}\sqrt{\frac{2a'_2+1}{2j_3+1}}\langle a'_2|\hat{V}^{\rm std}_v|a_2=j_3\rangle\,,\\
B(j_1,j_2,j_3;a''_2,a'_3)&:=C(j_1,j_2,j_3;a'_3)\times(2a''_2+1)(-1)^{a'_3-j_2+j_1+j_3}\begin{Bmatrix}
a'_3 & j_3 & a''_2\\
j_2 & j_1 & j_1
\end{Bmatrix}\,,\\
D(j_1,j_2,j_3;j'_2;a_2)&:=\sum_{j'_3}V(j'_3,j_1,j_2)(-1)^{2j_3}(2j'_3+1)(2j'_2+1)(2a_2+1)(-1)^{j_2+j'_2+\frac12}(-1)^{\frac12+j_1-j_2-a_2}(-1)^{1-2j_3}\notag\\
&\qquad\times
\begin{Bmatrix}
\frac12 & j_3 & a_2\\
j_1 & j'_2 & j_2
\end{Bmatrix}
\begin{Bmatrix}
\frac12 & j'_3 & j_3\\
\frac12 & a_2 & j_3
\end{Bmatrix}V(j_1,j'_2,j_3,1/2)\,,\\
E(j_1,j_2,j_3;j''_2;a'_2):=&\sum_{j'_1,a_3}V(j'_1,j_2,j_3)(-1)^{2j_3}(2j'_1+1)(2j''_2+1)(2a_3+1)(-1)^{j_2+j''_2+\frac12}(-1)^{j_1+j''_2-j_3+\frac12}(-1)^{1-2j_1}
\notag\\
&\qquad\times\begin{Bmatrix}
\frac12 & j_1 & a_3\\
j_3 & j''_2 & j_2
\end{Bmatrix}
\begin{Bmatrix}
\frac12 & j'_1 & j_1\\
\frac12 & a_3 & j_1
\end{Bmatrix}(2a'_2+1)(-1)^{\frac12+j_1-a_3}(-1)^{\frac12-j''_2+j_1+j_3}
\begin{Bmatrix}
j''_2 & j_1 & a'_2\\
\frac12 & j_3 & a_3
\end{Bmatrix}\,.
\end{align}
\end{widetext}
Here
\begin{align}
&C(j_1,j_2,j_3;a'_3):=\sum_{j'_1}V(j'_1,j_2,j_3)(-1)^{2j_3}(2j'_1+1)(2a'_3+1)\notag\\
&\hspace{3cm}\times(-1)^{\frac12+j_1-j'_1-a'_3}
\begin{Bmatrix}
\frac12 & \frac12 & a'_3\\
j_1 & j_1 & j'_1
\end{Bmatrix}\,,\\
&V(j'_1,j_2,j_3):=\frac{\ell_{\rm p}^3\,\beta^{3/2}}{4\sqrt{2}}\left[(j'_1+j_2+j_3+\frac32)(j'_1+j_2-j_3+\frac12)\right.\notag\\
&\hspace{1.5cm}\times\left.(j'_1-j_2+j_3+\frac12)(-j'_1+j_2+j_3+\frac12)\right]^{\frac14}\,,\\
&V(j_1,j'_2,j_3,1/2):=\frac{\ell_{\rm p}^3\,\beta^{3/2}}{4\sqrt{2}}\left|(j_1+j'_2+j_3+\frac32)\right.\notag\\
&\hspace{0.3cm}\times\left.(-j_1+j'_2+j_3+\frac12)(j_1-j'_2+j_3+\frac12)(j_1+j'_2-j_3+\frac12)\right|^\frac12
\end{align}
are all real numbers. Notice that the matrix elements $\langle a'_2|\hat{V}^{\rm std}_v|a_2=j_3\rangle$ in \eqref{A-volume-matrix} are all real (see Appendixes \ref{matrix-elements} and \ref{volume-eigenvalue} for proof) and the spin-network functions are orthogonal to each other. Hence the inner products between the spin-network functions in Eqs. \eqref{first-term} and \eqref{second-term} are also real. Thus we have completed the proof that the alternative volume operator acts trivially at the gauge-invariant trivalent vertices.

\section{New alternative inverse volume operator}\label{II}
To illuminate the meaning of an inverse volume operator, we recall that the Hamiltonian of a massless scalar field on a given spacetime is given by
\begin{align}\label{H-phi-all}
H_\phi(N)&=\frac12\int_\Sigma {\rm d}^3x\,N(x)\left[\frac{\pi^2}{\sqrt{\det(q)}}+\sqrt{\det(q)}\,q^{ab}(\partial_a\phi)\partial_b\phi\right](x)\notag\\
&\equiv\frac12\left[H_{{\rm kin},\phi}(N)+H_{{\rm der},\phi}(N)\right]\,,
\end{align}
where $\det(q)$ is the determinant of the spatial metric $q_{ab}$, and $\pi$ is the conjugate momentum of the scalar field. Consider the term $H_{{\rm kin},\phi}(N)$. In order to quantize it, we need to first regulate it. We can take the following fourfold point-splitting regularization for $H_{{\rm kin},\phi}(N)$,
\begin{widetext}
\begin{align}\label{H-phi-reg}
H_{{\rm kin},\phi}(N)&=\int_{\Sigma}{\rm d}^3x\,N(x)\frac{\pi^2(x)}{\sqrt{\det(q)(x)}}\notag\\
&=\lim_{\epsilon\rightarrow0}\int_{\Sigma}{\rm d}^3x\,N(x)\,\pi(x)\int_{\Sigma}{\rm d}^3y\,\pi(y)\int_{\Sigma}{\rm d}^3u\frac{\left|\det(e^i_a)\right|}{\left[\epsilon^3\sqrt{\det(q)}\right]^{3/2}}(u)\int_{\Sigma}{\rm d}^3w\frac{\left|\det(e^l_d)\right|}{\left[\epsilon^3\sqrt{\det(q)}\right]^{3/2}}(w)\;\chi_\epsilon(x,y)\chi_\epsilon(x,u)\chi_\epsilon(x,w)\notag\\
&=2^6\lim_{\epsilon\rightarrow0}\int_{\Sigma}{\rm d}^3x\,N(x)\,\pi(x)\int_{\Sigma}{\rm d}^3y\,\pi(y)\int_{\Sigma}{\rm d}^3u\;\left|-\frac{4}{3!}\tilde{\epsilon}^{abc}{\rm tr}\left({}^{(\frac12)}\!e_a{}^{(\frac12)}\!e_b{}^{(\frac12)}\!e_c\right)\right|\int_{\Sigma}{\rm d}^3w\;\left|-\frac{4}{3!}\tilde{\epsilon}^{def}{\rm tr}\left({}^{(\frac12)}\!e_d{}^{(\frac12)}\!e_e{}^{(\frac12)}\!e_f\right)\right|\notag\\
&\qquad\qquad\times\;\chi_\epsilon(x,y)\chi_\epsilon(x,u)\chi_\epsilon(x,w)\,,
\end{align}
\end{widetext}
where $\chi_\epsilon(x,y)$ denotes a characteristic function satisfying $\lim_{\epsilon\rightarrow0}\chi_\epsilon(x,y)/\epsilon^3=\delta^3(x,y)$; in the second step we have inserted $1=[\det(e^i_a)]^2/\left[\sqrt{\det(q)}\right]^2$\,; in the third step we used the identity $e_a(u)=\frac{2}{\kappa\beta}\{A_a(u),V(u,\epsilon)\}$ and absorbed $V(u,\epsilon)^{1/2}:=\epsilon^3\sqrt{\det(q)}(u)$ in the denominator into the Poisson bracket to get
 ${}^{(\frac12)}\!e_a(u):=\frac{2}{\kappa\beta}\{A_a(u),V(u,\epsilon)^{\frac12}\}$.

To quantize the first two integrals in \eqref{H-phi-reg}, we can easily replace $\pi$ by $-i\hbar\kappa\delta/\delta\phi$. It is clear that the only difference between the last two integrals in \eqref{H-phi-reg} and the expression \eqref{altv-classical} of $V(R)$ is the different powers of $V(x,\epsilon)$ in $e_a(x)$ and ${}^{(\frac12)}\!e_a(x)$. Thus we can also directly write down the quantum operators corresponding to the last two integrals. Hence we obtain the quantum version of $H_{{\rm kin},\phi}$ as
\begin{widetext}
\begin{align}
\hat{H}_{{\rm kin},\phi}(N)\cdot f_\gamma&=2^6(-i\hbar\kappa)^2\lim_{\epsilon\rightarrow0}\sum_{v'',v'''\in V(\gamma)}N(v'')X(v'')X(v''')\sum_{v\in V(\gamma)}\left|-\frac{4\times8}{3!E(v)}\sum_{s_I\cap s_J\cap s_K=v}\epsilon^{IJK}{\rm tr}\left({}^{(\frac12)}\!\hat{e}_I{}^{(\frac12)}\!\hat{e}_J{}^{(\frac12)}\!\hat{e}_K\right)\right|\notag\\
&\quad\times\sum_{v'\in V(\gamma)}\left|-\frac{4\times8}{3!E(v)}\sum_{s_L\cap s_M\cap s_N=v'}\epsilon^{LMN}{\rm tr}\left({}^{(\frac12)}\!\hat{e}_L{}^{(\frac12)}\!\hat{e}_M{}^{(\frac12)}\!\hat{e}_N\right)\right|\;\;\chi_\epsilon(v'',v''')\chi_\epsilon(v'',v)\chi_\epsilon(v'',v')\cdot f_\gamma\,,
\end{align}
\end{widetext}
where $X(v):=\frac12\left[X_R(v)+X_L(v)\right]$ is the sum over left and right invariant vector fields \cite{Thiemann:1997rt}, and
\begin{align}
{}^{(\frac12)}\!\hat{e}_I:= -\frac{2}{i\beta\ell_{\rm p}^2}h_{s_I}\left[h_{s_I}^{-1},\hat{V}^\frac12_{\rm std}\right]\,.
\end{align}
For small enough $\epsilon$, the nontrivial result corresponds to $v=v'=v''=v'''$. After taking the limit $\epsilon\rightarrow0$, we have
\begin{align}\label{H-phi-triad}
\hat{H}_{{\rm kin},\phi}(N)\cdot f_\gamma&=2^6(-i\hbar\kappa)^2\sum_{v\in V(\gamma)}N(v)X(v)X(v)\widehat{V^{-1}}^{\rm alt}_v\cdot f_\gamma\,,
 \end{align}
 where the new inverse volume operator is defined by
 \begin{align}\label{alt-inv-volum-op}
\widehat{V^{-1}}^{\rm alt}_v&:=\left|-\frac{4\times8}{3!E(v)}\sum_{s_I\cap s_J\cap s_K=v}\epsilon^{IJK}{\rm tr}\left({}^{(\frac12)}\!\hat{e}_I{}^{(\frac12)}\!\hat{e}_J{}^{(\frac12)}\!\hat{e}_K\right)\right|\notag\\
&\quad\times\left|-\frac{4\times8}{3!E(v)}\sum_{s_L\cap s_M\cap s_N=v}\epsilon^{LMN}{\rm tr}\left({}^{(\frac12)}\!\hat{e}_L{}^{(\frac12)}\!\hat{e}_M{}^{(\frac12)}\!\hat{e}_N\right)\right|
 \end{align}
which is the quantum version of $\frac{1}{\epsilon^3\sqrt{\det(q)}(x)}=\frac{1}{V(x,\epsilon)}$.  Note that the operator  ${}^{(\frac12)}\!\hat{e}_I$ differs from $\hat{e}_I$ only in the powers of $\hat{V}_{\rm std}$. Hence it is easy to see that the new alternative inverse volume $\widehat{V^{-1}}^{\rm alt}_v$ has the same qualitative properties as those of the alternative volume $\hat{V}^{\rm alt}_v$ in \eqref{alt-vol}. Thus $\widehat{V^{-1}}^{\rm alt}_v$ acts trivially at the gauge-invariant trivalent vertices. On the other hand, $\widehat{V^{-1}}^{\rm alt}_v$ also acts trivially at the coplanar vertices. Therefore $\hat{H}_{{\rm kin},\phi}(N)$ has trivial action at the coplanar vertices. This property is crucial in order to ensure that the full quantum constraint algebra is closed for the whole system of the matter coupled to gravity as pointed out in \cite{Thiemann:1996aw}.

\section{Summary and discussion}
In LQG, the cotriad operator plays an important role in the construction of the Hamiltonian constraint operators. As a consistency check, the fundamental flux operator can be reconstructed by the cotriad operator. The result of this paper shows that not only the flux operator but also a new volume operator \eqref{alt-vol} can be constructed by the cotriad operator. As shown in Sec. \ref{II-B}, the alternative volume operator \eqref{alt-vol} is internal gauge invariant, diffeomorphism covariant and symmetric. Its action on spin-network states vanishes at coplanar vertices as well as gauge-invariant trivalent vertices. For a nontrivial action, the operator \eqref{alt-vol} leaves the spins associated to the edges invariant but changes the intertwiners associated to the vertices. Thus the operator \eqref{alt-vol} shares the same qualitative properties with the standard volume operator in LQG. The successful construction of the alternative volume operator enhances the confidence in employing the cotriad operator as a building block in LQG.

There are two main differences between the standard volume operator \eqref{std-volume-def} and the new volume operator \eqref{alt-vol}. On one hand, the operational ways of the two operators on spin-network states are different from each other, although both results change the intertwiners of spin-network states. The standard one acts on spin-network states by the right-invariant vector fields $J^i_{e_I}$ on the holonomies along edges $e_I$ and hence changes the intertwiners $\vec{i}$ associated to vertices.  However, the new one is defined by the cotriad operator $\hat{e}_I:=-\frac{2}{i\beta\ell_{\rm p}^2}h_{s_I}\left[h_{s_I}^{-1},\hat{V}_{\rm std}\right]$. Hence its action on spin-network states first changes the spins $\vec{j}$ of edges by the action of holonomies in the intermediate steps,  and then changes the intertwiners $\vec{i}$ by the action of $\hat{V}_{\rm std}$. On the other hand, there is no ordering problem of the right-invariant vector field $J$'s in the standard volume operator \eqref{std-volume-def}, since the $J$'s associated to different edges commute with each other. However, the cotriad operators $\hat{e}_I$ associated to different edges do not commute with each other in general. Hence we have to choose an ordering of the three cotriad operators for a triple of edges incident at a vertex in the construction of the new volume operator \eqref{alt-vol}. Let us turn to the inverse volume operators. The essential constituents are certain cotriadlike operators.  Compared to the cotriadlike operator defined by ${}^{(\frac12)}\!\hat{e}^i_I:={\rm tr}\left(\tau_ih_{s_I}\left[h_{s_I}^{-1},\hat{V}^\frac12_{\rm std}\right]\right)$ in \cite{Brunnemann:2005ip}, the cotriadlike operator employed for the new inverse volume operator \eqref{alt-inv-volum-op} is defined by ${}^{(\frac12)}\!\hat{e}_I:=-\frac{2}{i\beta\ell_{\rm p}^2}h_{s_I}\left[h_{s_I}^{-1},\hat{V}^\frac12_{\rm std}\right]$. While the modification is slight, the operator ${}^{(\frac12)}\!\hat{e}_I$ is in more concise form and avoids the problem of the ordering of the Lie algebra elements $\tau_i$ in ${}^{(\frac12)}\!\hat{e}^i_I$. Moreover, our expression \eqref{alt-inv-volum-op} of the new inverse volume operator contains the absolute value to ensure its positivity, whereas the inverse volume operator in \cite{Brunnemann:2005ip} does not contain it. Since the new inverse volume operator takes the form similar to the new volume operator, it is convenient for us to compare the properties between the new inverse volume operator and the new volume operator.

In the loop quantization of gravity coupled to matter fields, one needs to employ an inverse volume operator to construct the matter Hamiltonian operator. According to the treatment of Eq. (4.1) in \cite{Brunnemann:2005ip} for the original inverse volume operator, one has to adopt the degenerate triangulation at coplanar vertices of spin networks in the regularization procedure of the Hamiltonian in order to obtain an on-shell anomaly-free quantum constraint algebra. It is shown in Sec. \ref{II} that a new inverse volume operator \eqref{alt-inv-volum-op} can be constructed in the light of the construction of the alternative volume operator. The expression of the alternative inverse volume operator \eqref{alt-inv-volum-op} implies that it is also possessed of the same qualitative properties of the alternative volume operator \eqref{alt-vol}. Thus the action of the operator \eqref{alt-inv-volum-op} on spin-network states vanishes at coplanar vertices, and hence the essential term \eqref{H-phi-triad} of the Hamiltonian operator for the scalar field also has trivial action on coplanar vertices. This property ensures that the quantum constraint algebra of the whole system of gravity coupled to the matter is anomaly free on shell without any special requirement on the regularization, though the action of the Hamiltonian constraint operator creates new coplanar vertices to the spin networks.

In the isotropic cosmological model, the inverse volume operator corresponds to $(1/a)^3$, where $a$ represents the scale factor. An interesting result in LQC is that the corresponding operator $\widehat{1/a}$ is bounded above and even vanishes on the zero eigenstate of the volume operator in LQC. This property is useful in understanding the big bang singularity resolution in LQC. While this property is not maintained by the inverse volume operator discussion in \cite{Brunnemann:2005ip}, the alternative inverse volume operator \eqref{alt-inv-volum-op} does maintain it. The reason is that we take an operators ordering in nature way, which is different from that in \cite{Brunnemann:2005ip}. Note that the cylindrical functions defined on the graphs with only trivalent or coplanar vertices are all zero eigenstates of the standard and alternative volume operators as well as the alternative inverse volume operator. Thus, the construction of the inverse volume operator \eqref{alt-inv-volum-op} opens a possible way to lift the result of singularity resolution of LQC to LQG.

\section*{ACKNOWLEDGMENTS}
J. Y thanks Chopin Soo and Hoi-Lai Yu for useful discussions. J. Y. is supported in part by NSFC Grant No. 11347006, by the Institute of Physics, Academia Sinica, Taiwan, and by the Natural Science Foundation of Guizhou University (Grant No. 47 in 2013). Y. M. is supported in part by the NSFC (Grants No. 11235003 and No. 11475023) and the Research Fund for the Doctoral Program of Higher Education of China.

\appendix
\renewcommand\thesubsection{\Alph{section}.\arabic{subsection}}
\renewcommand\theequation{\Alph{section}\arabic{equation}}

\section{Derivation of two matrix elements of the volume operator in Eq. \eqref{eigenvalue-matrix}}\label{matrix-elements}
In this appendix, we calculate two matrix elements of the volume operator in Eq. \eqref{eigenvalue-matrix} for $I=1,J=2,K=3$ using the graphical method introduced in \cite{graph-I,graph-II}.

\subsection{The first matrix element in Eq. \eqref{eigenvalue-matrix}}
The two states on which $\hat{V}^{\rm std}_v$ acts in the first term of \eqref{eigenvalue-matrix} take the form (see the derivation in \cite{graph-II})
\begin{align}\label{state-1-1}
&{\left(h_{s_3}\hat{V}^{\rm std}_vh_{s_3}^{-1}\right)^A}_C\cdot T^{v,s}_{\gamma,\vec{j},\vec{i}}(A)\notag\\
&=\sum_{j'_3}V(j'_3,j_1,j_2)(-1)^{2j_3}(2j'_3+1)\makeSymbol{
\includegraphics[width=4cm]{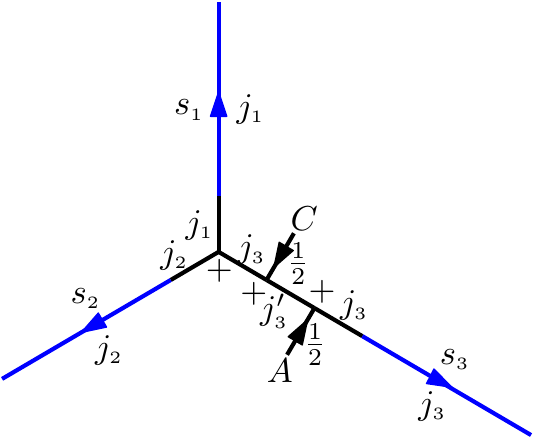}}\,,
\end{align}
and
\begin{align}
&{\left(h_{s_1}\hat{V}^{\rm std}_vh_{s_1}^{-1}\right)^A}_C\cdot T^{v,s}_{\gamma,\vec{j},\vec{i}}(A)\notag\\
&=\sum_{j'_1}V(j'_1,j_2,j_3)(-1)^{2j_3}(2j'_1+1)\makeSymbol{
\includegraphics[width=3.3cm]{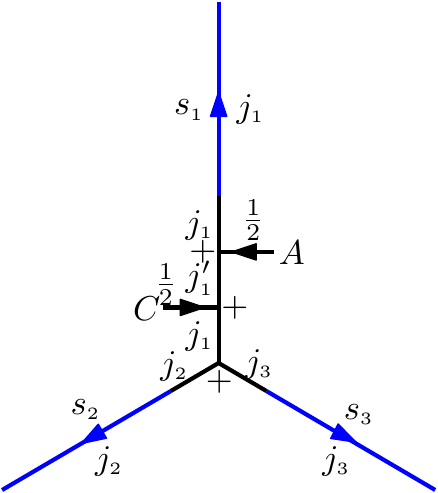}}\,,\label{state-1-2}
\end{align}
\begin{widetext}
where $V(j'_3,j_1,j_2)\equiv V(1/2,j'_3,j_1,j_2;a_2=j'_3+1/2,a_3=j_2), V(j'_1,j_2,j_3)\equiv V(1/2,j'_1,j_2,j_3;a_2=j'_1+1/2,a_3=j_3)$ with
\begin{align}
V(1/2,j'_1,j_2,j_3;a_2=j'_1+1/2,a_3=j_3)\equiv\frac{\ell_{\rm p}^3\,\beta^{3/2}}{4\sqrt{2}}\left[(j'_1+j_2+j_3+\frac32)(j'_1+j_2-j_3+\frac12)(j'_1-j_2+j_3+\frac12)(-j'_1+j_2+j_3+\frac12)\right]^{\frac14}\,.
\end{align}
\end{widetext}
Now, let us consider the action of $\hat{V}^{\rm std}_v$ on the state \eqref{state-1-1}. The intertwiner of the state \eqref{state-1-1}  can be transformed as
\begin{align}\label{intertwine-volume-1}
&\makeSymbol{
\includegraphics[width=1.4cm]{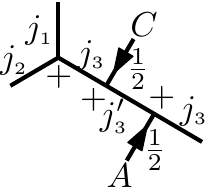}}=\makeSymbol{
\includegraphics[width=1.4cm]{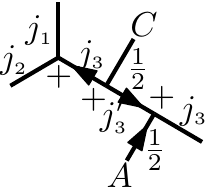}}\notag\\
&=\sum_{a_3\in \{0,1\}}(2a_3+1)(-1)^{\frac12+j_3-j'_3-a_3}
\begin{Bmatrix}
\frac12 & \frac12 & a_3\\
j_3 & j_3 & j'_3
\end{Bmatrix}\makeSymbol{
\includegraphics[width=1.6cm]{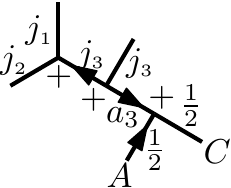}}\notag\\
&=\sum_{a_3\in \{0,1\}}(2a_3+1)(-1)^{\frac12-j_3-j'_3-a_3}
\begin{Bmatrix}
\frac12 & \frac12 & a_3\\
j_3 & j_3 & j'_3
\end{Bmatrix}\makeSymbol{
\includegraphics[width=1.6cm]{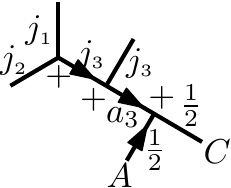}}\,,
\end{align}
where in the second step we used [Eq. (A.63) in \cite{graph-I}]
\begin{align}\label{6j-interchange-1}
\makeSymbol{
\includegraphics[width=2cm]{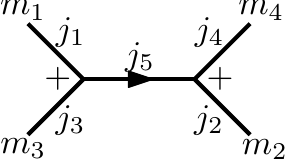}}&=\sum_{j_6}(2j_6+1)(-1)^{j_1+j_4-j_5-j_6}
\begin{Bmatrix}
j_1 & j_2 & j_6\\
j_4 & j_3 & j_5
\end{Bmatrix}\notag\\
&\quad\times
\makeSymbol{
\includegraphics[width=2cm]{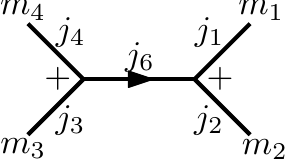}}\,.
\end{align}
Note that the volume operator only changes the intermediate momenta in the intertwiner. In the expression \eqref{intertwine-volume-1}, the open segments with spins $j_1,j_2,j_3$ are jointed with the segments $s_1,s_2,s_3$. Hence the volume operator only changes the intermediate momenta $a_2$ (between $j_1$ and $j_3$). The allowed intermediated momenta $a_2$ in \eqref{intertwine-volume-1} are determined by the values of $a_3$ and $j_3$. There are the following two cases:
\begin{widetext}
\begin{itemize}
\item $j_3=\frac12$
\begin{align}
a_3&=0\;\Rightarrow\; a_2=j_3\hspace{1.2cm}\Leftrightarrow\; \text{one-dimensional intertwiner space}\,,\notag\\
a_3&=1\;\Rightarrow\; a_2=
\begin{cases}
j_3\\
j_3+1
\end{cases}\;\Leftrightarrow\; \text{two-dimensional intertwiner space}\,;
\end{align}
\item $j_3\geqslant1$
\begin{align}
a_3&=0\;\Rightarrow\; a_2=j_3\hspace{1.2cm}\Leftrightarrow\; \text{one-dimensional intertwiner space}\,,\notag\\
a_3&=1\;\Rightarrow\; a_2=
\begin{cases}
j_3-1\\
j_3\\
j_3+1
\end{cases}\;\Leftrightarrow\; \text{three-dimensional intertwiner space}\,.
\end{align}
\end{itemize}
Hence the action of $\hat{V}^{\rm std}_v$ changes the intertwiner in \eqref{state-1-1} as
\begin{align}
\hat{V}^{\rm std}_v\makeSymbol{
\includegraphics[width=1.4cm]{graph/identity-1}}&=\sum_{a_3\in \{0,1\}}(2a_3+1)(-1)^{\frac12-j_3-j'_3-a_3}
\begin{Bmatrix}
\frac12 & \frac12 & a_3\\
j_3 & j_3 & j'_3
\end{Bmatrix}\hat{V}^{\rm std}_v\makeSymbol{
\includegraphics[width=1.6cm]{graph/identity-2}}\notag\\
&=3(-1)^{-\frac12-j_3-j'_3}\begin{Bmatrix}
\frac12 & \frac12 & 1\\
j_3 & j_3 & j'_3
\end{Bmatrix}\hat{V}^{\rm std}_v\makeSymbol{
\includegraphics[width=1.6cm]{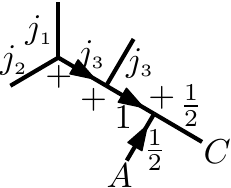}}\notag\\
&=3(-1)^{-\frac12-j_3-j'_3}\begin{Bmatrix}
\frac12 & \frac12 & 1\\
j_3 & j_3 & j'_3
\end{Bmatrix}\sum_{a'_2}\sqrt{\frac{2a'_2+1}{2j_3+1}}\langle a'_2|\hat{V}^{\rm std}_v|a_2=j_3\rangle\makeSymbol{
\includegraphics[width=1.6cm]{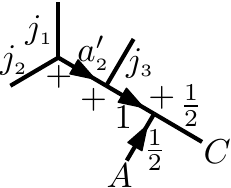}}\notag\\
&=3(-1)^{-\frac12-j_3-j'_3}\begin{Bmatrix}
\frac12 & \frac12 & 1\\
j_3 & j_3 & j'_3
\end{Bmatrix}\sum_{a'_2}\sqrt{\frac{2a'_2+1}{2j_3+1}}\langle a'_2|\hat{V}^{\rm std}_v|a_2=j_3\rangle(-1)^{1+j_3+a'_2}\makeSymbol{
\includegraphics[width=1.6cm]{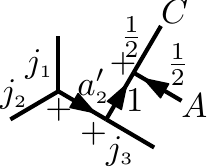}}\notag\\
&=\sum_{a'_2}3(-1)^{\frac12+a'_2-j'_3}\begin{Bmatrix}
\frac12 & \frac12 & 1\\
j_3 & j_3 & j'_3
\end{Bmatrix}\sqrt{\frac{2a'_2+1}{2j_3+1}}\langle a'_2|\hat{V}^{\rm std}_v|a_2=j_3\rangle\makeSymbol{
\includegraphics[width=1.6cm]{graph/identity-4-1}}\,,
\end{align}
\end{widetext}
where in the second step we used the result that the volume operator vanishes on the one-dimensional intertwiner space. A detailed calculation shows that the matrix elements $\langle a'_2|\hat{V}^{\rm std}_v|a_2=j_3\rangle$ are all real for $j_3\geqslant1$  (see Appendix \ref{volume-eigenvalue} for proof), and $\langle a'_2|\hat{V}^{\rm std}_v|a_2\rangle=\delta_{a'_2,a_2}\langle a_2=j_3|\hat{V}^{\rm std}_v|a_2=j_3=\frac12\rangle$ are also real for $j_3=\frac12$ since the the volume operator is automatically diagonal in two-dimensional intertwiner space (see Appendix A in \cite{graph-II} for proof). Hence we obtain the action of $\hat{V}^{\rm std}_v$ on the state \eqref{state-1-1} as
\begin{align}
&\hat{V}^{\rm std}_v\left|{\left(h_{s_3}\hat{V}^{\rm std}_vh_{s_3}^{-1}\right)^A}_C\cdot T^{v,s}_{\gamma,\vec{j},\vec{i}}(A)\right\rangle\notag\\
&=\sum_{a'_2}A(j_1,j_2,j_3;a'_2)\makeSymbol{
\includegraphics[width=3.8cm]{graph/graph-snf-6}}\label{state-1-3} \,,
\end{align}
where
\begin{align}
&A(j_1,j_2,j_3;a'_2):=\sum_{j'_3}V(j'_3,j_1,j_2)(-1)^{2j_3}(2j'_3+1)3(-1)^{\frac12+a'_2-j'_3}\notag\\
&\hspace{1.2cm}\times\begin{Bmatrix}
\frac12 & \frac12 & 1\\
j_3 & j_3 & j'_3
\end{Bmatrix}\sqrt{\frac{2a'_2+1}{2j_3+1}}\langle a'_2|\hat{V}^{\rm std}_v|a_2=j_3\rangle
\end{align}
are real numbers.

To calculate the action of $\hat{V}^{\rm std}_v$ on the state \eqref{state-1-2}, it is convenient to change the form of the state $\left|\left(h_{s_1}\hat{V}^{\rm std}_vh_{s_1}^{-1}\right)_{AC}\cdot T^{v,s}_{\gamma,\vec{j},\vec{i}}(A)\right\rangle$ into the one similar to Eq. \eqref{state-1-3}. Using the identity \eqref{6j-interchange-1}, Eq. \eqref{state-1-2} reduces to
\begin{align}\label{state-1-4}
&\left|{\left(h_{s_1}\hat{V}^{\rm std}_vh_{s_1}^{-1}\right)^A}_C\cdot T^{v,s}_{\gamma,\vec{j},\vec{i}}(A)\right\rangle\notag\\
&=\sum_{a'_3}C(j_1,j_2,j_3;a'_3)\makeSymbol{
\includegraphics[width=3.1cm]{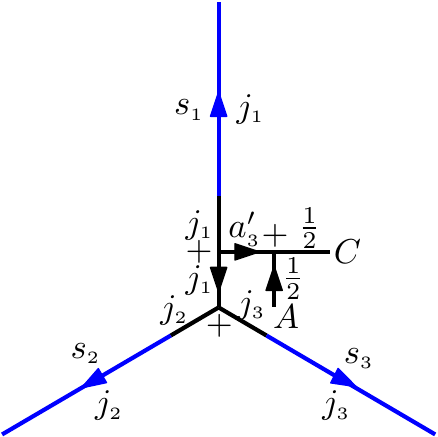}}\notag\\
&=\sum_{a'_3,a''_2}B(j_1,j_2,j_3;a''_2,a'_3)\makeSymbol{
\includegraphics[width=3.3cm]{graph/graph-snf-8}}\,,
\end{align}
where
\begin{align}
&C(j_1,j_2,j_3;a'_3):=\sum_{j'_1}V(j'_1,j_2,j_3)(-1)^{2j_3}(2j'_1+1)(2a'_3+1)\notag\\
&\hspace{3cm}\times(-1)^{\frac12+j_1-j'_1-a'_3}
\begin{Bmatrix}
\frac12 & \frac12 & a'_3\\
j_1 & j_1 & j'_1
\end{Bmatrix}
\end{align}
and
\begin{align}
&B(j_1,j_2,j_3;a''_2,a'_3):=C(j_1,j_2,j_3;a'_3)\times(2a''_2+1)\notag\\
&\hspace{3cm}\times(-1)^{a'_3-j_2+j_1+j_3}\begin{Bmatrix}
a'_3 & j_3 & a''_2\\
j_2 & j_1 & j_1
\end{Bmatrix}
\end{align}
are real numbers, and in the second step, we have used the identity [Eq. (A.64) in \cite{graph-I}]
\begin{align}\label{6j-interchange-2}
\makeSymbol{
\includegraphics[width=2cm]{graph/3j-6j-1}}&=\sum_{j_6}(2j_6+1)(-1)^{j_1-j_2+j_3+j_4}
\begin{Bmatrix}
j_1 & j_4 & j_6\\
j_2 & j_3 & j_5
\end{Bmatrix}\notag\\
&\qquad\times\makeSymbol{
\includegraphics[width=2cm]{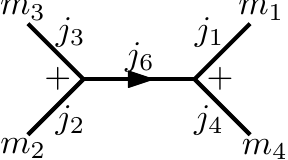}}\,.
\end{align}

By \eqref{state-1-3} and \eqref{state-1-4}, we obtain the first matrix element of the volume operator in Eq. \eqref{eigenvalue-matrix} as
\begin{align}
&\left\langle{\left(h_{s_1}\hat{V}^{\rm std}_vh_{s_1}^{-1}\right)^A}_C\cdot T^{v,s}_{\gamma,\vec{j},\vec{i}}(A)\right|\hat{V}^{\rm std}_v\left|{\left(h_{s_3}\hat{V}^{\rm std}_vh_{s_3}^{-1}\right)^A}_C\cdot T^{v,s}_{\gamma,\vec{j},\vec{i}}(A)\right\rangle\notag\\
=&\sum_{a'_2,a''_2,a'_3}A(j_1,j_2,j_3;a'_2)B(j_1,j_2,j_3;a''_2,a'_3)\notag\\
&\times\left\langle\makeSymbol{
\includegraphics[width=3.3cm]{graph/graph-snf-8}}\right|\left.\makeSymbol{
\includegraphics[width=3.3cm]{graph/graph-snf-6}}\right\rangle\,.
\end{align}

\subsection{The second matrix element in Eq. \eqref{eigenvalue-matrix}}
The state acted on by $\hat{V}^{\rm std}_v$ in the second term of \eqref{eigenvalue-matrix} takes the form
\begin{widetext}
\begin{align}
{\left(h_{s_2}^{-1}h_{s_3}\hat{V}^{\rm std}_vh_{s_3}^{-1}\right)^B}_C\cdot T^{v,s}_{\gamma,\vec{j},\vec{i}}(A)
&=\sum_A{(h_{s_2}^{-1})^B}_A{\left(h_{s_3}\hat{V}^{\rm std}_vh_{s_3}^{-1}\right)^A}_C\left[(-1)^{2j_3}\makeSymbol{
\includegraphics[width=3cm]{graph/graph-snf-1}}\right]\notag\\
&=\sum_A{(h_{s_2}^{-1})^B}_A\sum_{j'_3}V(j'_3,j_1,j_2)(-1)^{2j_3}(2j'_3+1)\makeSymbol{
\includegraphics[width=4cm]{graph/graph-snf-2}}\notag\\
&=\sum_{j'_3}V(j'_3,j_1,j_2)(-1)^{2j_3}(2j'_3+1)\sum_{j'_2}(2j'_2+1)(-1)^{j_2+j'_2+\frac12}\sum_A\makeSymbol{
\includegraphics[width=4.4cm]{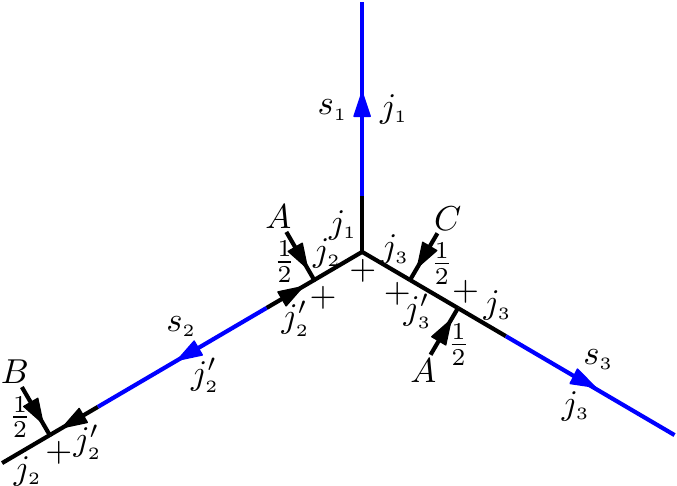}}\notag\\
&=\sum_{j'_3}V(j'_3,j_1,j_2)(-1)^{2j_3}(2j'_3+1)\sum_{j'_2}(2j'_2+1)(-1)^{j_2+j'_2+\frac12}\notag\\
&\quad\times\sum_{a_2}(2a_2+1)(-1)^{\frac12+j_1-j_2-a_2}(-1)^{1-2j_3}
\begin{Bmatrix}
\frac12 & j_3 & a_2\\
j_1 & j'_2 & j_2
\end{Bmatrix}
\begin{Bmatrix}
\frac12 & j'_3 & j_3\\
\frac12 & a_2 & j_3
\end{Bmatrix}\makeSymbol{
\includegraphics[width=4cm]{graph/graph-snf-5}}\,,
\end{align}
where in the last step we have used
\begin{align}
\sum_A\makeSymbol{
\includegraphics[width=1.6cm]{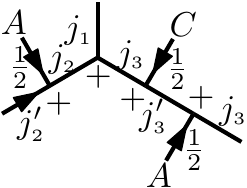}}&=\sum_A\makeSymbol{
\includegraphics[width=1.6cm]{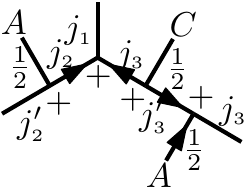}}=(-1)^{2j_3}\sum_A\makeSymbol{
\includegraphics[width=1.6cm]{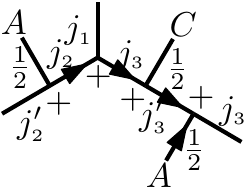}}=(-1)^{2j_3}\sum_{a_2}(2a_2+1)(-1)^{\frac12+j_1-j_2-a_2}
\begin{Bmatrix}
\frac12 & j_3 & a_2\\
j_1 & j'_2 & j_2
\end{Bmatrix}
\sum_A\makeSymbol{
\includegraphics[width=1.8cm]{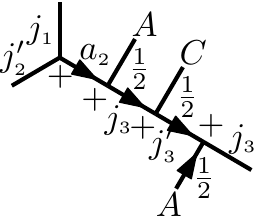}}\notag\\
&=(-1)^{2j_3}\sum_{a_2}(2a_2+1)(-1)^{\frac12+j_1-j_2-a_2}
\begin{Bmatrix}
\frac12 & j_3 & a_2\\
j_1 & j'_2 & j_2
\end{Bmatrix}\sum_{a_3}(2a_3+1)(-1)^{\frac12+\frac12-j_3-a_3}
\begin{Bmatrix}
\frac12 & j'_3 & a_3\\
\frac12 & a_2 & j_3
\end{Bmatrix}
\sum_A\makeSymbol{
\includegraphics[width=1.8cm]{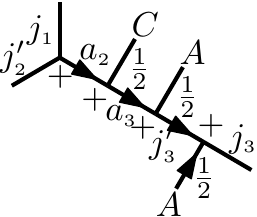}}\notag\\
&=(-1)^{2j_3}\sum_{a_2}(2a_2+1)(-1)^{\frac12+j_1-j_2-a_2}
\begin{Bmatrix}
\frac12 & j_3 & a_2\\
j_1 & j'_2 & j_2
\end{Bmatrix}\sum_{a_3}(2a_3+1)(-1)^{\frac12+\frac12-j_3-a_3}
\begin{Bmatrix}
\frac12 & j'_3 & a_3\\
\frac12 & a_2 & j_3
\end{Bmatrix}
\makeSymbol{
\includegraphics[width=1.8cm]{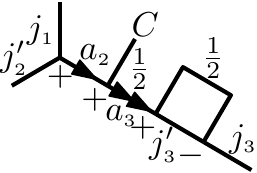}}\notag\\
&=(-1)^{2j_3}\sum_{a_2}(2a_2+1)(-1)^{\frac12+j_1-j_2-a_2}
\begin{Bmatrix}
\frac12 & j_3 & a_2\\
j_1 & j'_2 & j_2
\end{Bmatrix}\sum_{a_3}(2a_3+1)(-1)^{\frac12+\frac12-j_3-a_3}
\begin{Bmatrix}
\frac12 & j'_3 & a_3\\
\frac12 & a_2 & j_3
\end{Bmatrix}
\frac{(-1)^{2j_3}\delta_{a_3,j_3}}{2j_3+1}\makeSymbol{
\includegraphics[width=1.2cm]{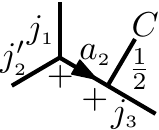}}\notag\\
&=\sum_{a_2}(2a_2+1)(-1)^{\frac12+j_1-j_2-a_2}(-1)^{1-2j_3}
\begin{Bmatrix}
\frac12 & j_3 & a_2\\
j_1 & j'_2 & j_2
\end{Bmatrix}
\begin{Bmatrix}
\frac12 & j'_3 & j_3\\
\frac12 & a_2 & j_3
\end{Bmatrix}
\makeSymbol{
\includegraphics[width=1.2cm]{graph/identity-11}}\,.
\end{align}
Here we have used the identity \eqref{6j-interchange-1} in the third and fourth steps. Notice that the intertwiner space is two dimensional. Hence the volume operator is diagonalizable as
\begin{align}
\hat{V}^{\rm std}_v\makeSymbol{
\includegraphics[width=1.2cm]{graph/identity-11}}&=-\frac{1}{\sqrt{2a_2+1}}\hat{V}^{\rm std}_v\left[\sqrt{2(2a_2+1)}\makeSymbol{
\includegraphics[width=4.5cm]{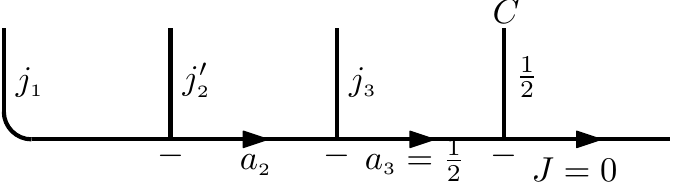}}\right]\notag\\
&=-\frac{1}{\sqrt{2a_2+1}}V(j_1,j'_2,j_3,1/2)\left[\sqrt{2(2a_2+1)}\makeSymbol{
\includegraphics[width=4.5cm]{graph/identity-12}}\right]\notag\\
&=V(j_1,j'_2,j_3,1/2)\makeSymbol{
\includegraphics[width=1.2cm]{graph/identity-11}}\,,
\end{align}
where
\begin{align}
V(j_1,j'_2,j_3,1/2)&\equiv V(j_1,j'_2,j_3,\frac12;a_2=j_3+\frac12,a_3=\frac12)=\frac{\ell_{\rm p}^3\,\beta^{3/2}}{4\sqrt{2}}\left|\left\langle a_2-1=j_3-\frac12\right|\hat{q}_{123}\left|a_2=j_3+\frac12\right\rangle\right|^\frac12\notag\\
&=\frac{\ell_{\rm p}^3\,\beta^{3/2}}{4\sqrt{2}}\left|(j_1+j'_2+j_3+\frac32)(-j_1+j'_2+j_3+\frac12)(j_1-j'_2+j_3+\frac12)(j_1+j'_2-j_3+\frac12)\right|^\frac12\,.
\end{align}
Hence we have
\begin{align}\label{state-1-5}
\hat{V}^{\rm std}_v\left|{\left(h_{s_2}^{-1}h_{s_3}\hat{V}^{\rm std}_vh_{s_3}^{-1}\right)^B}_C\cdot T^{v,s}_{\gamma,\vec{j},\vec{i}}(A)\right\rangle
&=\sum_{j'_2,a_2}D(j_1,j_2,j_3;j'_2;a_2)\makeSymbol{
\includegraphics[width=4.4cm]{graph/graph-snf-5}}\,,
\end{align}
where
\begin{align}
D(j_1,j_2,j_3;j'_2;a_2)&:=\sum_{j'_3}V(j'_3,j_1,j_2)(-1)^{2j_3}(2j'_3+1)(2j'_2+1)(2a_2+1)(-1)^{j_2+j'_2+\frac12}(-1)^{\frac12+j_1-j_2-a_2}(-1)^{1-2j_3}\notag\\
&\qquad\times
\begin{Bmatrix}
\frac12 & j_3 & a_2\\
j_1 & j'_2 & j_2
\end{Bmatrix}
\begin{Bmatrix}
\frac12 & j'_3 & j_3\\
\frac12 & a_2 & j_3
\end{Bmatrix}V(j_1,j'_2,j_3,1/2)
\end{align}
is a real number. On the other hand, the state on the left of $\hat{V}^{\rm std}_v$ in the second term of \eqref{eigenvalue-matrix} can be simplified as
\begin{align}
&{\left(h_{s_2}^{-1}h_{s_1}\hat{V}^{\rm std}_vh_{s_1}^{-1}\right)^B}_C\cdot T^{v,s}_{\gamma,\vec{j},\vec{i}}(A)\notag\\
&=\sum_{j'_1}V(j'_1,j_2,j_3)(-1)^{2j_3}(2j'_1+1)\sum_{j''_2}(2j''_2+1)(-1)^{j_2+j''_2+\frac12}\sum_{a_3}(2a_3+1)(-1)^{j_1+j''_2-j_3+\frac12}(-1)^{1-2j_1}
\begin{Bmatrix}
\frac12 & j_1 & a_3\\
j_3 & j''_2 & j_2
\end{Bmatrix}
\begin{Bmatrix}
\frac12 & j'_1 & j_1\\
\frac12 & a_3 & j_1
\end{Bmatrix}\notag\\
&\qquad\times\makeSymbol{
\includegraphics[width=4.4cm]{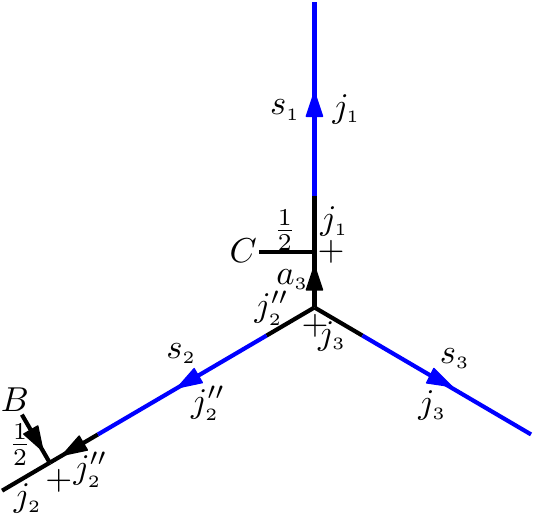}}\notag\\
&=\sum_{j'_1}V(j'_1,j_2,j_3)(-1)^{2j_3}(2j'_1+1)\sum_{j''_2}(2j''_2+1)(-1)^{j_2+j''_2+\frac12}\sum_{a_3}(2a_3+1)(-1)^{j_1+j''_2-j_3+\frac12}(-1)^{1-2j_1}
\begin{Bmatrix}
\frac12 & j_1 & a_3\\
j_3 & j''_2 & j_2
\end{Bmatrix}
\begin{Bmatrix}
\frac12 & j'_1 & j_1\\
\frac12 & a_3 & j_1
\end{Bmatrix}\notag\\
&\quad\times\sum_{a'_2}(2a'_2+1)(-1)^{\frac12+j_1-a_3}(-1)^{\frac12-j''_2+j_1+j_3}
\begin{Bmatrix}
j''_2 & j_1 & a'_2\\
\frac12 & j_3 & a_3
\end{Bmatrix}\makeSymbol{
\includegraphics[width=4.4cm]{graph/graph-snf-10}}\,.
\end{align}
Hence, we have
\begin{align}\label{state-1-6}
{\left(h_{s_2}^{-1}h_{s_1}\hat{V}^{\rm std}_vh_{s_1}^{-1}\right)^B}_C\cdot T^{v,s}_{\gamma,\vec{j},\vec{i}}(A)
&=\sum_{j''_2,a'_2}E(j_1,j_2,j_3;j''_2;a'_2)\makeSymbol{
\includegraphics[width=4.4cm]{graph/graph-snf-10}}\,,
\end{align}
where
\begin{align}
E(j_1,j_2,j_3;j''_2;a'_2):=&\sum_{j'_1,a_3}V(j'_1,j_2,j_3)(-1)^{2j_3}(2j'_1+1)(2j''_2+1)(2a_3+1)(-1)^{j_2+j''_2+\frac12}(-1)^{j_1+j''_2-j_3+\frac12}(-1)^{1-2j_1}
\notag\\
&\qquad\times\begin{Bmatrix}
\frac12 & j_1 & a_3\\
j_3 & j''_2 & j_2
\end{Bmatrix}
\begin{Bmatrix}
\frac12 & j'_1 & j_1\\
\frac12 & a_3 & j_1
\end{Bmatrix}(2a'_2+1)(-1)^{\frac12+j_1-a_3}(-1)^{\frac12-j''_2+j_1+j_3}
\begin{Bmatrix}
j''_2 & j_1 & a'_2\\
\frac12 & j_3 & a_3
\end{Bmatrix}\,.
\end{align}

Thus, by \eqref{state-1-5}  and \eqref{state-1-6} we obtain the second matrix element of the volume operator in Eq. \eqref{eigenvalue-matrix} as
\begin{align}
&\left\langle{\left(h_{s_2}^{-1}h_{s_1}\hat{V}^{\rm std}_vh_{s_1}^{-1}\right)^B}_C\cdot T^{v,s}_{\gamma,\vec{j},\vec{i}}(A)\right|\hat{V}^{\rm std}_v\left|{\left(h_{s_2}^{-1}h_{s_3}\hat{V}^{\rm std}_vh_{s_3}^{-1}\right)^B}_C\cdot T^{v,s}_{\gamma,\vec{j},\vec{i}}(A)\right\rangle\notag\\
=&\sum_{j'_2,j''_2,a_2,a'_2}D(j_1,j_2,j_3;j'_2;a_2)E(j_1,j_2,j_3;j''_2;a'_2)\left\langle\makeSymbol{
\includegraphics[width=4.4cm]{graph/graph-snf-10}}\right|\left.\makeSymbol{
\includegraphics[width=4.4cm]{graph/graph-snf-5}}\right\rangle\,.
\end{align}
\end{widetext}

\section{The action of the volume operator on the three-dimensional intertwiner space}\label{volume-eigenvalue}
Let us denote the three states in the intertwiner space associated to $v$ by
\begin{align}
|\alpha_1\rangle&\equiv|a_2=j_3-1\rangle\,,\\
|\alpha_2\rangle&\equiv|a_2=j_3\rangle\,,\\
|\alpha_3\rangle&\equiv|a_2=j_3+1\,\rangle\,.
\end{align}
Then the matrix of the operator $i\hat{q}_{123}$ reads
\begin{align}\label{q-1-2-3-matrix}
(i\hat{q}_{123})&=
\begin{pmatrix}
\langle\alpha_1|i\hat{q}_{123}|\alpha_1\rangle & \langle\alpha_1|i\hat{q}_{123}|\alpha_2\rangle & \langle\alpha_1|i\hat{q}_{123}|\alpha_3\rangle\\
\langle\alpha_2|i\hat{q}_{123}|\alpha_1\rangle & \langle\alpha_2|i\hat{q}_{123}|\alpha_2\rangle & \langle\alpha_2|i\hat{q}_{123}|\alpha_3\rangle\\
\langle\alpha_3|i\hat{q}_{123}|\alpha_1\rangle & \langle\alpha_3|i\hat{q}_{123}|\alpha_2\rangle & \langle\alpha_3|i\hat{q}_{123}|\alpha_3\rangle
\end{pmatrix}\notag\\
&=:\begin{pmatrix}
0 & ia & -ib\\
-ia & 0 & -ic\\
ib & ic & 0
\end{pmatrix}\,,
\end{align}
where $a,b,c$, as the matrix elements of $\hat{q}_{123}$, are real numbers. The eigenvalues and corresponding (orthonormal) eigenvectors of $i\hat{q}_{123}$ are given by
\begin{align}
\lambda_1&=0\rightarrow |\vec{e}_1\rangle=\frac{c}{\Delta}\begin{pmatrix}
1\\
-\frac{b}{c}\\
-\frac{a}{c}
\end{pmatrix},\notag\\
\lambda_2&=\Delta\rightarrow |\vec{e}_2\rangle=\frac{\sqrt{b^2+c^2}}{\sqrt{2}\,\Delta}
\begin{pmatrix}
\frac{ac-ib\Delta}{b^2+c^2}\\
-\frac{ab+ic\Delta}{b^2+c^2}\\
1
\end{pmatrix},\notag\\
\lambda_3&=-\Delta\rightarrow |\vec{e}_3\rangle=\frac{\sqrt{b^2+c^2}}{\sqrt{2}\,\Delta}
\begin{pmatrix}
\frac{ac+ib\Delta}{b^2+c^2}\\
-\frac{ab-ic\Delta}{b^2+c^2}\\
1
\end{pmatrix}\,,
\end{align}
where $\Delta\equiv\sqrt{a^2+b^2+c^2}$. Hence we get
\begin{align}
\hat{V}^{\rm std}_v|\alpha_2\rangle&=\frac{\ell_{\rm p}^3\,\beta^{3/2}}{4\sqrt{2}}\sqrt{|i\hat{q}_{123}|}\;|\alpha_2\rangle\notag\\
&=\frac{\ell_{\rm p}^3\,\beta^{3/2}}{4\sqrt{2}}\sqrt{|i\hat{q}_{123}|}\;\left(\sum_{i=1,2,3}|\vec{e}_i\rangle\langle\vec{e}_i|\right)|\alpha_2\rangle\notag\\
&=\frac{\ell_{\rm p}^3\,\beta^{3/2}}{4\sqrt{2}}\sum_{i=1,2,3}\sqrt{|i\hat{q}_{123}|}\;|\vec{e}_i\rangle\langle\vec{e}_i|\alpha_2\rangle\notag\\
&=\frac{\ell_{\rm p}^3\,\beta^{3/2}}{4\sqrt{2}}\left(\sqrt{|\lambda_2|}|\vec{e}_2\rangle\langle\vec{e}_2|\alpha_2\rangle+\sqrt{|\lambda_3|}|\vec{e}_3\rangle\langle\vec{e}_3|\alpha_2\rangle\right)\notag\\
&=\frac{\ell_{\rm p}^3\,\beta^{3/2}}{4\sqrt{2}}\sqrt{\Delta}\left(|\vec{e}_2\rangle\langle\vec{e}_2|\alpha_2\rangle+|\vec{e}_3\rangle\langle\vec{e}_3|\alpha_2\rangle\right)\notag\\
&=\frac{\ell_{\rm p}^3\,\beta^{3/2}}{4\sqrt{2}}\frac{1}{\Delta^{3/2}}\left[bc|\alpha_1\rangle+(a^2+c^2)|\alpha_2\rangle-ab|\alpha_3\rangle\right]\,,
\end{align}
which shows that $\hat{V}^{\rm std}_v$ changes $|\alpha_2\rangle$ into a linear composition of $|\alpha_i\rangle$ with real factors.
Hence the matrix elements $\langle a'_2|\hat{V}^{\rm std}_v|a_2=j_3\rangle$ are also real.

%----------------------------------------------------
\providecommand{\href}[2]{#2}\begingroup\raggedright\endgroup

%----------------------------------------------------

\end{document}